\documentclass[final,3p,times,onecolumn]{elsarticle}
\pdfoutput=1
\usepackage{amsmath,amssymb,bm,graphicx,bbold,epsf,colordvi}
\usepackage{lipsum}
\usepackage{braket}
\allowdisplaybreaks 
\usepackage{color}
\newcommand\f{\frac}
\newcommand\as{\alpha_s}
\newcommand{\ba}{\begin{eqnarray}}
\newcommand{\ea}{\end{eqnarray}}
\newcommand{\bea}{\begin{eqnarray}}
\newcommand{\eea}{\end{eqnarray}}
\newcommand{\bef}{\begin{figure}[t]\centering}
\newcommand{\eef}{\end{figure}}
\newcommand{\be}{\begin{equation}}
\newcommand{\ee}{\end{equation}}
\newcommand{\nn}{\nonumber}

\def\OMIT#1{{}}

\begin{document}

\title{Soft drop groomed jet angularities at the LHC}

\author{Zhong-Bo Kang\fnref{label1,label2,label3}}
\ead{zkang@physics.ucla.edu}
\author{Kyle Lee\fnref{label4,label5}}
\ead{kunsu.lee@stonybrook.edu}
\author{Xiaohui Liu\fnref{label6}}
\ead{xiliu@bnu.edu.cn}
\author{Felix Ringer\fnref{label7}}
\ead{fmringer@lbl.gov}
\address[label1]{Department of Physics and Astronomy, University of California, Los Angeles, CA 90095, USA}
\address[label2]{Mani L. Bhaumik Institute for Theoretical Physics, University of California, Los Angeles, CA 90095, USA}
\address[label3]{Theoretical Division, Los Alamos National Laboratory, Los Alamos, NM 87545, USA}
\address[label4]{C.N. Yang Institute for Theoretical Physics, Stony Brook University, Stony Brook, NY 11794, USA}
\address[label5]{Department of Physics and Astronomy, Stony Brook University, Stony Brook, NY 11794, USA}
\address[label6]{Center of Advanced Quantum Studies, Department of Physics, Beijing Normal University, Beijing 100875, China}
\address[label7]{Nuclear Science Division, Lawrence Berkeley National Laboratory, Berkeley, CA 94720, USA}


\begin{abstract}
Jet angularities are a class of jet substructure observables where a continuous parameter is introduced in order to interpolate between different classic observables such as the jet mass and jet broadening. We consider jet angularities measured on an inclusive jet sample at the LHC where the soft drop grooming procedure is applied in order to remove soft contaminations from the jets. The soft drop algorithm allows for a precise comparison between theory and data and could be used to extract the QCD strong coupling constant $\alpha_s$ from jet substructure data in the future. We develop a framework to realize the resummation of all relevant large logarithms at the next-to-leading logarithmic (NLL) accuracy. To demonstrate that the developed formalism is suitable for the extraction of $\alpha_s$, we extend our calculations to next-to-next-to-leading logarithm (NNLL) for the jet mass case. Overall, we find good agreement between our NLL numerical results and Pythia simulations for LHC kinematics and we observe an improved agreement when the NNLL components are included. In addition, we expect that groomed jet angularities will be a useful handle for studying the modification of jets in heavy-ion collisions. 
\end{abstract}

\maketitle

\section{Introduction \label{sec:intro}}

In recent years significant progress has been made in achieving a quantitative understanding of jet substructure observables. Jet substructure techniques allow for a wide range of applications, see~\cite{Larkoski:2017jix,Asquith:2018igt} for recent reviews. Observables like the jet mass distribution have a large non-perturbative (NP) contribution at the LHC making a comparison of data with purely perturbative results in QCD problematic. However, the NP contribution can be systematically reduced by making use of grooming algorithms that are designed to remove soft wide angle radiation from the observed jet which then also need to be taken into account in the perturbative calculations. The grooming procedure discussed in this work is soft drop declustering~\cite{Larkoski:2014wba} which has several advantages from a theoretical point of view~\cite{Dasgupta:2013ihk}. The soft drop groomed jet mass distribution was calculated in~\cite{Frye:2016aiz,Marzani:2017mva,Kang:2018jwa} and the corresponding LHC measurements from ATLAS and CMS can be found in~\cite{Aaboud:2017qwh,Sirunyan:2018xdh}. A good agreement between theory and experimental data was obtained for the shape of the jet mass distribution. In this work, we extend the groomed jet mass calculation of~\cite{Kang:2018jwa} to a more general class of observables which are known as (groomed) jet angularities $\tau_a$ with the following definition~\cite{Berger:2003iw,Almeida:2008yp,Gras:2017jty}
\be\label{eq:jetangularity}
\tau_a = \f{1}{p_T}\sum_{i\in J}p_{Ti} \,\Delta R_{iJ}^{2-a} \, .
\ee
Here the $p_{Ti}$ denote the transverse momenta of the particles $i$ in the jet and $\Delta R_{iJ}^2=(\Delta\eta_{iJ})^2+(\Delta\phi_{iJ})^2$ is their distance to the jet axis. The sum runs over all particles in the (groomed) jet and $p_T$ in the denominator is the jet's transverse momentum. Here $a$ is a free parameter that controls the sensitivity to collinear radiation and it smoothly interpolates between different traditional jet shape variables. Note that part of the existing literature adopted a different convention $2-a\equiv\alpha$ for the exponent of $\Delta R_{iJ}$ in Eq.~(\ref{eq:jetangularity}). We consider the cross section
\be\label{eq:jetangularitycrsec}
\f{1}{\sigma_{\rm incl}}\f{d\sigma}{d\eta dp_Td\tau_a} \,,
\ee
differential in the jet angularity variable $\tau_a$ and the observed jet's transverse momentum $p_T$ and rapidity $\eta$. We consider inclusive jet production $pp\to{\rm jet}+X$ and thus normalize by the total inclusive jet production cross section $\sigma_{\rm incl}$. Jet angularities without grooming have been discussed in the literature, see for example~\cite{Ellis:2010rwa,Hornig:2016ahz,Kang:2018qra}. In this work, we focus on the impact of the soft drop grooming procedure on the jet angularity measurements. The soft drop grooming procedure can be summarized as follows. First, the observed jet is reclustered with the Cambridge/Aachen (C/A) algorithm~\cite{Dokshitzer:1997in,Wobisch:1998wt}. Second, the soft drop criterion 
\be\label{eq:sd}
\f{{\rm min}[p_{T1},p_{T2}]}{p_{T1}+p_{T2}} > z_{\rm cut}\left(\f{\Delta R_{12}}{R}\right)^\beta \,.
\ee
is checked at each clustering step when going backward through the C/A clustering tree. Here $p_{T1,2}$ are the momenta of the two branches that had been merged together and $\Delta R_{12}$ is their geometric distance in the $\eta$-$\phi$ plane. When the softer branch fails the criterion it is removed from the jet and the procedure continues until it is satisfied. The remaining particles in the jet constitute the soft drop groomed jet. A convenient choice can be made for the soft threshold parameter $z_{\rm cut}$ and the angular exponent $\beta$. For $\beta=0$ the soft drop procedure reduces to the modified mass drop tagger (mMDT) developed in~\cite{Dasgupta:2013ihk}. The jet angularity measurement is performed only on the particles that remain in the groomed jet when the soft drop procedure ends.

We see two important applications for the observables discussed here which we outline in the following. First, as it was recently proposed in~\cite{Bendavid:2018nar}, the QCD strong coupling constant $\alpha_s$ can be determined from groomed jet substructure observables. Similarly, event shape variables in $e^+e^-$ collisions have been established as important benchmark processes to constrain the strong coupling constant. See~\cite{Abbate:2010xh,Hoang:2015hka,Baron:2018nfz,Banfi:2018mcq,Procura:2018zpn,Bell:2018gce} for analyses along those lines. Groomed jet angularities or also energy-energy correlation functions~\cite{Banfi:2004yd} constitute a natural extension from event shapes in $e^+e^-$ collisions to the more complicated environment in $pp$ collisions at the LHC. 

From the theoretical side, a precise extraction of the strong coupling constant requires a sophisticated understanding of the jet angularities at fixed order including resummation and NP effects for $\tau_a\to 0$. In this work, we perform the resummation of all relevant logarithms as specified in the next section at next-to-leading logarithmic (NLL) accuracy. For the jet mass case, $a=0$ in Eq.~(\ref{eq:jetangularity}), with $\beta=0$ we also extend the resummation to next-to-next-to-leading logarithmic (NNLL) order. The desired precision for a competitive extraction of $\as$ is next-to-next-to-leading order (NNLO) supplemented with the resummation at NNLL accuracy. This accuracy is achieved for the $e^+e^-$ event shape variables mentioned above that are included in the determination of the world average of the QCD strong coupling constant $\alpha_s(M_Z^2)= 0.1181\pm 0.0011$~\cite{Tanabashi:2018oca}. In this work we focus specifically on the jet substructure of an inclusive jet sample $pp\to{\rm jet}+X$~\cite{Kaufmann:2015hma,Kang:2016ehg,Dai:2016hzf} which can be extended systematically beyond the currently achieved precision in the future. Also note that we consistently normalize the jet angularity cross section in Eq.~(\ref{eq:jetangularitycrsec}) to the inclusive jet production cross section $\sigma_{\rm incl}$. Such a normalization is desired for the extraction of $\as$ and the inclusive jet production cross section is under good theoretical control after dedicated theoretical efforts in the past years. In~\cite{Currie:2016bfm}, the full NNLO calculation was completed and in~\cite{Liu:2017pbb,Liu:2018ktv} the joint resummation of threshold and jet radius logarithms was carried out. Therefore, the normalization of the cross section chosen in Eq.~(\ref{eq:jetangularitycrsec}) appears as a natural choice and the perturbative accuracy can be extended systematically in the future. See also~\cite{Kardos:2018kth}. In addition, inclusive jet cross sections can be measured with the highest experimental statistics.

Secondly, we expect that (groomed) jet angularities can be a useful tool for jet studies in heavy-ion collisions. In~\cite{Acharya:2017goa,Sirunyan:2018gct,ATLAS:2018jsv}, it was found that in heavy-ion collisions both the groomed and ungroomed jet mass distributions are unmodified relative to the $pp$ baseline within the experimental uncertainties. However, other jet substructure observables show a significant modification due to the presence of the QCD medium. For example, the closely related jet broadening or girth, $a=1$ in Eq.~(\ref{eq:jetangularity}), was measured by ALICE in~\cite{Acharya:2018uvf} which exhibits a large non-trivial modification pattern. By measuring jet angularities in heavy-ion collisions for different values of $a$ and $\beta$ it will be possible to systematically map out which jet substructure observables are modified and it will help to understand the underlying dynamics. Our work thus provides another step toward utilizing jets as precision probes of the quark-gluon plasma.

The remainder of this paper is organized as follows. In section~\ref{sec:2}, we outline the factorization formalism developed in this work. In section~\ref{sec:3}, we present numerical results and compare to Pythia simulations for exemplary LHC kinematics. We draw our conclusions in section~\ref{sec:4} and we present an outlook.

\section{Theoretical framework \label{sec:2}}

In the first part of this section, we introduce the factorization theorem used in this work for the soft drop groomed jet angularities. We present the relevant functions and their associated renormalization group (RG) evolution equations at NLL accuracy. In the second part, we extend the framework to NNLL for the jet mass case with $\beta=0$.

\subsection{Groomed jet angularities}

In this section we outline the factorization structure for the groomed jet angularity cross section. Throughout this work we use the framework of Soft Collinear Effective Theory (SCET)~\cite{Bauer:2000ew, Bauer:2000yr, Bauer:2001ct, Bauer:2001yt,Beneke:2002ph} and we follow the framework for inclusive jet production $pp\to\text{jet}+X$ developed in~\cite{Kang:2018qra,Kang:2018jwa}. For sufficiently narrow jets with a radius $R\ll 1$~\cite{Kaufmann:2015hma,Kang:2016mcy,Dai:2016hzf} we can write the triple differential cross section in Eq.~(\ref{eq:jetangularitycrsec}) as
\ba\label{eq:factorization}
\hspace*{-0.6cm}\f{d\sigma}{d\eta dp_Td\tau_a}\,&=&\sum_{abc}f_a(x_a,\mu)\otimes f_b(x_b,\mu)\otimes H_{ab}^c(x_a,x_b,\eta, p_T/z,\mu)\otimes {\cal G}_c(z,p_TR,\tau_a,\mu,z_{\rm cut},\beta)\,.
\ea
Here $f_{a,b}$ denote the parton distribution functions to find partons $a,b$ in the colliding protons. The parton's momentum fractions are denoted by $x_{a,b}$. The hard functions $H_{ab}^c$ describe the hard-scattering event $ab\to c$ to produce a final state parton which has a transverse momentum of $p_T/z$ that fragments into the observed jet. The production of the jet is described by the semi-inclusive jet functions ${\cal G}_c$. They depend on the transverse momentum fraction $z$ contained in the jet relative to that of the initial parton and, in addition, ${\cal G}_c$ captures the information about the angularity $\tau_a$ of the observed jet. In addition, it depends on the soft drop grooming parameters $z_{\rm cut},\beta$ following Eq.~(\ref{eq:sd}). Similar to parton-to-hadron fragmentation functions, the semi-inclusive jet functions ${\cal G}_c$ satisfy RG equations which take the form of standard DGLAP evolution equations
\ba\label{eq:dglap}
\mu\f{d}{d\mu}{\cal G}_i(z,p_T R,\tau_a,\mu,z_{\rm cut},\beta)\,&= &\sum_j P_{ji}(z) \otimes\, {\cal G}_j(z,p_T R,\tau_a,\mu,z_{\rm cut},\beta) \,.
\ea
Here, $P_{ji}(z)$ denote the Altarelli-Parisi splitting kernels. By solving the DGLAP evolution equations between the scales $\mu\sim p_T R\to p_T$, the resummation of single logarithms in the jet radius parameter $\alpha_s^n\ln^n R$ can be achieved at NLL following the numerical procedure of~\cite{Vogt:2004ns,Bodwin:2015iua}. Power corrections $\sim{\cal O}(R^2)$ to Eq.~(\ref{eq:factorization}) are generally expected to be small, even for a relatively large jet radius $R\sim 1$, see for example Refs~\cite{Kang:2018jwa,Mukherjee:2012uz}. On the other hand, the advantages of working in the small-$R$ limit are for example the possibility to define quark-gluon fractions perturbatively order-by-order in QCD through Eq.~(\ref{eq:factorization}). In addition, it is possible to include non-global logarithms~\cite{Dasgupta:2001sh} in $z_{\rm cut}$~\cite{Kang:2018jwa} and it is possible to study universality aspects of the relevant nonperturbative physics, see~\cite{Korchemsky:1999kt,Lee:2006nr,Stewart:2014nna}. In particular the universality aspect of the nonperturbative corrections is an essential ingredient for the extraction of $\alpha_s$ which is often fitted simultaneously with a nonperturbative shape function, see~\cite{Abbate:2010xh,Hoang:2015hka}, and which is discussed in more detail in section~\ref{sec:3}. In the phenomenologically relevant kinematic regime with the scaling 
\be\label{eq:scaling}
\tau_a^{1/(2-a)}/R\ll z_{\rm cut}\ll 1 \,,
\ee
large logarithms may spoil the convergence of the perturbative expansion of the cross section which requires the resummation to all orders. This can be achieved by a refactorization of the semi-inclusive jet function ${\cal G}_c$. Each function then obeys it's own RG evolution equation which eventually allows for the all order resummation of the relevant large logarithms. We find
\ba\label{eq:refactorization}
\hspace*{-0.5cm}{\cal G}_c(z,p_T R,\tau_a,\mu,z_{\rm cut},\beta)\,&=&\sum_i {\cal H}_{c\to i}(z,p_T R,\mu) S^{\notin{\rm gr}}_i(z_{\rm cut}p_T R,\beta,\mu) \nn\\
 &&\times \int d\tau_a^{C_i} d\tau_a^{S_i}\,\delta(\tau_a-\tau_a^{C_i}-\tau_a^{S_i})C_i(\tau_a^{C_i},p_T,\mu)\, S_i^{\rm gr}(\tau_a^{S_i},p_T,R,\mu,z_{\rm cut},\beta) \,.
\ea
The hard matching functions ${\cal H}_{c\to i}$ describe how a parton $c$ coming from the hard interaction initiates a jet with parton $i$. They take into account energetic radiation at the scale $\mu\sim p_T R$ outside of the observed jet~\cite{Kang:2017mda,Kang:2017glf} as they are not allowed to contribute to the observed jet angularity with the scaling in Eq.~(\ref{eq:scaling}). The soft functions $S_i^{\notin {\rm gr}}$ take into account soft radiation that always fail the soft drop criterion in Eq.~(\ref{eq:sd}). Therefore, $S_i^{\notin {\rm gr}}$ does not depend on the observed jet angularity $\tau_a$. On the other hand, $S_i^{\rm gr}$ takes into account soft radiation boosted along the direction of the jet that may or may not pass the soft drop criterion which introduces the dependence on $\tau_a$ and $z_{\rm cut},\beta$. The remaining collinear mode $C_i$~\cite{Hornig:2009vb} takes into account collinear radiation in the jet, which parametrically always passes the soft drop criterion and thus it is insensitive to $z_{\rm cut},\beta$ and it contributes to the observed jet angularity $\tau_a$. The radiation associated with the collinear function is sufficiently energetic such that it is not affected by the grooming algorithm up to power corrections. In addition, the collinear mode does not probe the jet boundary and therefore the collinear function is independent of $R$. See~\cite{Frye:2016aiz,Kang:2018jwa} for further discussions of the obtained factorization structure.

The refactorized form of the semi-inclusive jet function in Eq.~(\ref{eq:refactorization}) can be derived analogously to the groomed jet mass case, $a=0$~\cite{Kang:2018jwa}. Note that each function in Eq.~(\ref{eq:refactorization}) depends only on a single scale which allows for the resummation of all relevant large logarithms. Within the effective field theory framework this is achieved by evaluating each function at its natural scale which eliminates the large logarithms at fixed order. Using RG evolution techniques all functions can then be evolved to a common scale through which the all order resummation is achieved. Here, for practical reasons, typically the scale $\mu\sim p_T$ of the hard functions in Eq.~(\ref{eq:factorization}) is used. Besides the resummation of single logarithms in the jet radius parameter $R$, which is achieved by solving Eq.~(\ref{eq:dglap}), we perform the NLL resummation of double logarithms in $\alpha_s^n\ln^{2n}(\tau_a^{1/(2-a)}/R)$ and $\alpha_s^n\ln^{2n}z_{\rm cut}$. Note that for all relevant functions in Eq.~(\ref{eq:refactorization}) it is possible to write down definitions at the operator level. We refer the interested reader to~\cite{Kang:2018jwa} where the operator definitions for the soft drop jet mass distribution were given within SCET.

\begin{figure*}[t]
\includegraphics[width=\textwidth]{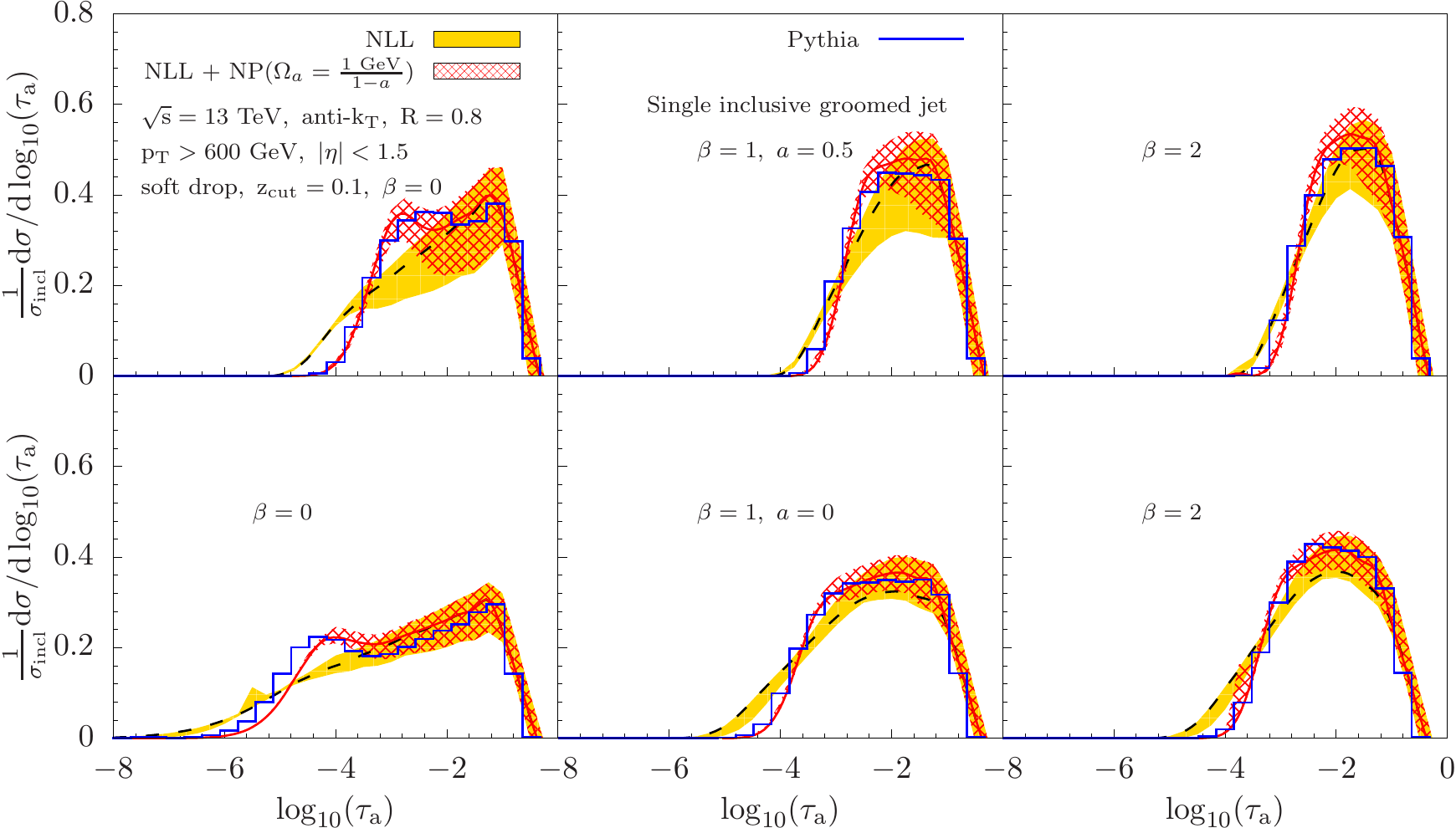}
\caption{
Groomed jet angularities at the LHC for different values of $a=0.5$ (upper panels) and $a=0$ (lower panels), $z_{\rm cut}=0.1$ and $\beta=0,1,2$ (from left to right). We choose $\sqrt{s}=13$~TeV, $p_T>600$~GeV, $|\eta|<1.5$ for an inclusive jet sample reconstructed using the anti-k$_T$ algorithm. The purely perturbative result at NLL is shown (yellow band)  as well as results using Pythia (blue). In addition, we show the perturbative result after including non-perturbative effects using the shape function as discussed in the text (hatched red band).
\label{fig:angularity1}
}
\end{figure*}

We are now going to present the results for the soft function $S_i^{\rm gr}$ that takes into account soft radiation boosted along the jet and may pass the soft drop criterion and thus contribute to the observed jet angularity. The bare result at NLO is given by 
\ba\label{eq:soft}
\hspace*{-0.8cm}S^{\rm gr}_i(\tau_a,p_T,R,\mu,z_{\rm cut},\beta)\, & = &\, \delta(\tau_a)+\f{\as}{\pi}\f{C_i}{1-a} \left[\f{2-a+\beta}{1+\beta}\left(-\f{1}{\epsilon^2}+\f{\pi^2}{24}\right)\delta(\tau_a)+\f{A}{\epsilon}\left(\f{1}{A\tau_a}\right)_+  - \f{2(1+\beta)}{2-a+\beta}A\left(\f{\ln (A\tau_a)}{A\tau_a}\right)_+ \right]\,,
\ea
where $C_i=C_{F,A}$ for $i=q,g$ and the variable $A$ is given by
\be
A=\Bigg(\left(\f{z_{\rm cut}}{R^\beta}\right)^{\f{1-a}{2-a+\beta}}\f{p_T}{\mu}\Bigg)^{\f{2-a+\beta}{1+\beta}} \,.
\ee
There are two limiting cases that can be checked for consistency. For $a=0$, Eq.~(\ref{eq:soft}) reduces to the soft function of the groomed jet mass distribution~\cite{Kang:2018jwa} and for $\beta\to\infty$, we obtain the ungroomed jet angularity soft function of~\cite{Ellis:2010rwa,Kang:2018qra}. After performing the renormalization of the bare soft functions $S^{\rm gr}_i$ in Eq.~(\ref{eq:soft}), the RG evolution equations are obtained as
\ba
\mu\f{d}{d\mu}S_i^{\rm gr}(\tau_a,p_T,R,\mu,z_{\rm cut},\beta)\,&= &\int d\tau_a' \gamma_{S_i}^{\rm gr}(\tau_a-\tau_a',p_T,R,\mu,z_{\rm cut},\beta) S_i^{\rm gr}(\tau_a',p_T,R,\mu,z_{\rm cut},\beta)\,, \nn\\ 
\ea
where the anomalous dimensions $\gamma^{\rm gr}_{S_i}$ are given by
\ba
\gamma_{S_i}^{\rm gr}(\tau_a,p_T,R,\mu,z_{\rm cut},\beta)\,&=&\,\f{\as}{\pi}\f{C_i}{1-a}\left[\left(\f{1}{\tau_a}\right)_++\f{2-a+\beta}{2+2\beta}\ln(A)\delta(\tau_a)\right]  \,.
\ea
Note that the collinear functions $C_i$ and the soft functions $S_i^{\notin {\rm gr}}$ satisfy similar evolution equations which can be found together with the RG equations for the hard matching functions ${\cal H}_{c\to i}$ in~\cite{Kang:2018jwa}. Besides the hard scale $\mu\sim p_T$ and the jet scale $\mu_J\sim p_T R$, we summarize here for completeness the natural scales of the collinear function and the two soft functions
\ba\label{eq:scales}
&&\mu_C\sim p_T \tau_a^{1/(2-a)}\,,\qquad \mu_S^{\notin {\rm gr}}\sim z_{\rm cut} p_T R \,, \qquad \mu_S^{\rm gr}\sim p_T \left(\f{z_{\rm cut}}{R^\beta}\right)^{\f{1-a}{2-a+\beta}} \tau_a^{\f{1+\beta}{2-a+\beta}} \,.
\ea
We note that there is a transition point of the groomed jet angularity distribution to the ungroomed case at large values of $\tau_a$. After this transition point the radiation is sufficiently energetic that grooming does not play a role anymore. The soft function $S_i^{\rm gr}$ reduces to the ungroomed jet angularity soft function of~\cite{Ellis:2010rwa,Kang:2018qra} and, in addition, the soft function taking into account radiation that fails soft drop reduces to unity $S_i^{\notin {\rm gr}}\to 1$. This can be seen explicitly by analyzing the phase space constraints for the soft function $S_i^{\rm gr}$. Both the jet algorithm and the soft drop grooming criterion put constraints on the soft radiation. Above the jet angularity value of
\be
\tau_a=z_{\rm cut}R^{2-a} \,,
\ee
for $\beta-a>-2$, the soft drop constraint is less restrictive than the jet algorithm constraint and it can be dropped. Thus the soft function reduces to its ungroomed analogue. Note that the transition point here depends only on $a$ but it is independent of the parameter $\beta$. This transition point is found at NLO but also holds for the evolution in the sense that the two soft scales merge yielding the ungroomed soft scale $\mu_S^{\rm ungr}$ of~\cite{Kang:2018qra},
\be
\mu_{{\rm gr}}|_{\tau_a=z_{\rm cut}R^{2-a}}=\mu_{\notin {\rm gr}}|_{\tau_a=z_{\rm cut}R^{2-a}}=\f{p_T\tau_a}{R^{1-a}}=\mu_S^{\rm ungr}\,.
\ee
In addition, the anomalous dimensions for the evolution of the two soft functions add up to the ungroomed case~\cite{Kang:2018jwa}
\be\label{eq:con}
\gamma_{S_i}^{\notin{\rm gr}}\,\delta(\tau_a)+\gamma^{\rm gr}_{S_i}=\gamma^{\rm ungr}_{S_i} \,.
\ee
This transition point can also be seen from our numerical results presented in the next section. Independent of $\beta$ the numerical results intersect at this point, exceptt for a numerically small remaining dependence on $\beta$ due to the fixed order expressions of the soft functions.

\subsection{The groomed jet mass distribution at NNLL}

\begin{figure*}[t]
\includegraphics[width=\textwidth]{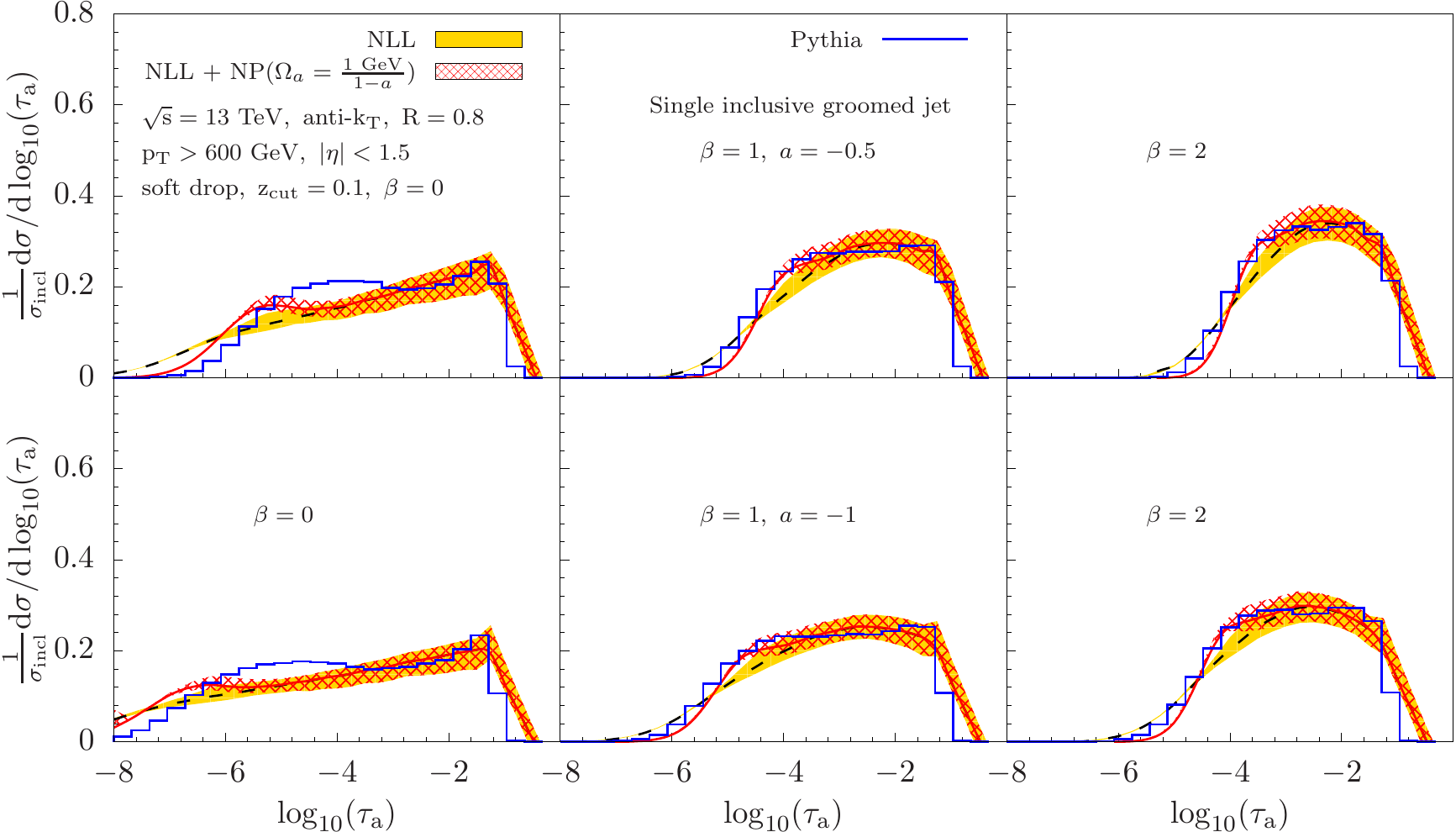}
\caption{
Same as Fig.~\ref{fig:angularity1} but for $a=-0.5$ (upper panels) and $a=-1$ (lower panels).
\label{fig:angularity2}
}
\end{figure*}

In order to perform the resummation at NNLL order for the groomed jet mass distribution with $\beta=0$ we need the anomalous dimensions of all the relevant functions in Eq.~(\ref{eq:refactorization}) beyond one-loop. For the collinear function $C_i$ and the two soft functions $S_i^{\notin{\rm gr}}$ and $S_i^{\rm gr}$, we can generally write the evolution equations as
\be
\frac{dF(\mu)}{d \ln \mu} 
= \left[ 2\,\Gamma_{\rm cusp}[\alpha_s] \ln \frac{\mu}{Q} + \gamma[\alpha_s] \right] F(\mu) \,.
\ee
The multiplicative form of the RG equations holds for $S_i^{\notin{\rm gr}}$ and also for $C_i$ and $S_i^{\rm gr}$ after taking the Fourier transform of Eq.~(\ref{eq:refactorization}). The cusp $\Gamma_{\rm cusp}[\as]$ and the non-cusp contribution $\gamma[\as]$ can be written as a perturbative expansion in the strong coupling constant $\alpha_s$ as
\be
\Gamma_{\rm cusp}=\sum_{n=0}^\infty \left(\f{\as}{4\pi}\right)^{n+1} \Gamma_n \,, \qquad \gamma=\sum_{n=0}^\infty \left(\f{\as}{4\pi}\right)^{n+1} \gamma^{(n)} \,.
\ee
To achieve NNLL accuracy, we include the cusp anomalous dimension up to three-loop order~\cite{Korchemsky:1987wg,Moch:2004pa}. In addition, we need the two-loop non-cusp contributions $\gamma^{(1)}$. The relevant results for the collinear function can be obtained from~\cite{Becher:2006qw,Becher:2010pd}. 
The anomalous dimension $\gamma^{\rm gr}_{S_i}$ for the collinear soft function is also universal and its NNLO result can be found in~\cite{Frye:2016aiz}, which is
\ba
\gamma^{{\rm gr},(1)}_{S_i} &=& C_i \left[
-17.005 \,C_F + \left(
-55.20 + \frac{22\pi^2}{9} + 56 \zeta_3
\right) \, C_A  + \left(
23.61 - \frac{8\pi^2}{9}
\right) n_f T_R
\right] \,,
\ea
where $C_i = C_F$ for quark jets and $C_i = C_A$ for gluon jets.

To extract the two-loop non-cusp anomalous dimension $\gamma_{S_i}^{\notin{\rm gr},(1)}$ for the soft function $S_i^{\notin{\rm gr}}$, we utilize the fact that 
the sum of the anomalous dimensions of the two groomed soft functions has to be equal to the anomalous dimension of the soft function that appears in the ungroomed calculation, see Eq.~(\ref{eq:con}). We explicitly performed the calculation of the two-loop non-cusp anomalous dimension $\gamma^{\rm ungr,(1)}_{S_i}$ for the soft function in the ungroomed case by relating the ungroomed jet mass to the hemisphere mass~\cite{Kelley:2011ng}. We isolated the non-global contribution and we checked our results against the ones in~\cite{Chien:2015cka} to find full agreement. Using the consistency relation in Eq.~(\ref{eq:con}), we find 
\ba
\gamma_{S_i}^{\notin{\rm gr},(1)}&=& C_i \left[
17.005\, C_F
+ \left( 
25.2741 - \frac{11}{9}\pi^2 - 28\zeta_3 
\right)C_A +
\left( -15.3137 + \frac{4}{9}\pi^2 \right) T_Rn_f
\right] \,.
\ea
It is interesting to note that $\gamma_{S_i}^{\notin{\rm gr}}$ is found to be numerically identical to half of the anomalous dimension of the global soft function $S_{\rm G}$ in~\cite{Frye:2016aiz} up to two-loop. The relative numerical error after taking into account the factor of 1/2  is of the order of ${\cal O}(10^{-5})$. 

We thus have all the relevant ingredients to perform the resummation at NNLL accuracy up to non-global logarithms~\cite{Dasgupta:2001sh}. Their contribution is expected to be small as they only appear as logarithms of $z_{\rm cut} = 0.1 $~\cite{Kang:2018jwa}.

\section{Phenomenology \label{sec:3}}

In this section we present numerical results using the theoretical formalism presented above. In addition, we compare to Pythia8 results~\cite{Sjostrand:2014zea} for LHC kinematics and we analyze the structure of the NP contribution to the cross section. The scale $\mu_S^{\rm gr}$ of the soft function $S_i^{\rm gr}$ is the lowest scale in our calculation. As it approaches $\Lambda_{\rm QCD}$, NP effects start to become important. By identifying $\mu_S^{\rm gr}\simeq\Lambda_{\rm QCD}$, we find that NP effects are relevant in the region
\be
\tau_a\simeq \left(\frac{\Lambda_{\rm QCD}R^\beta}{p_T z_{\rm cut}}\right)^{\f{1-a}{1+\beta}} \f{\Lambda_{\rm QCD}}{p_T} \,.
\ee
Following~\cite{Korchemsky:1999kt}, we include NP effects using a shape function which is convolved with the purely perturbative result in Eq.~(\ref{eq:factorization}). The argument $\tau_a$ of the purely perturbative cross section is shifted by the virtuality of the soft mode $S_i^{\rm gr}$~\cite{Frye:2016aiz}. We thus have
\ba\label{eq:convolutionNP}
\hspace*{-0.5cm}\f{d\sigma}{d\eta dp_T d\tau_a}\, &=&\int dk\, F(k) \f{d\sigma^{\rm pert}}{d\eta dp_Td\tau_a}\left(\tau_a - \left(\f{k R^\beta}{p_T z_{\rm cut}}\right)^{\f{1-a}{1+\beta}}\f{k}{p_T}\right) \,.
\ea
We adopt the following model for the NP shape function~\cite{Stewart:2014nna}
\be\label{eq:Fk}
F(k)=\f{4k}{\Omega_a^2}\exp(-2k/\Omega_a)\,,
\ee
which is normalized to unity and it only depends on a single parameter $\Omega_a$ which is given by its first moment. Alternatively, it is also common to use for example Pythia to estimate the non-perturbative contribution. In addition, we use profile scales~\cite{Ligeti:2008ac} in order to smoothly freeze the relevant scales in Eq.~(\ref{eq:scales}) above the Landau pole. Throughout this work we use the CT14 PDF set of~\cite{Dulat:2015mca}. We choose to fix the scale $\mu_S^{\notin {\rm gr}}$ relative to the jet scale $\mu_J$ and, in addition, we relate the collinear scale $\mu_C$ to $\mu_S^{\rm gr}$, see Eq.~(\ref{eq:scales}). The QCD scale uncertainty bands presented in this section are obtained by varying all scales by factors of 2 around their canonical choices and by taking the envelope. We do not include the uncertainties from the NP model and the 
$\alpha_s$ prescription used to deal with the Landau pole. 

\begin{figure}[t]
\centering
\includegraphics[width=10cm]{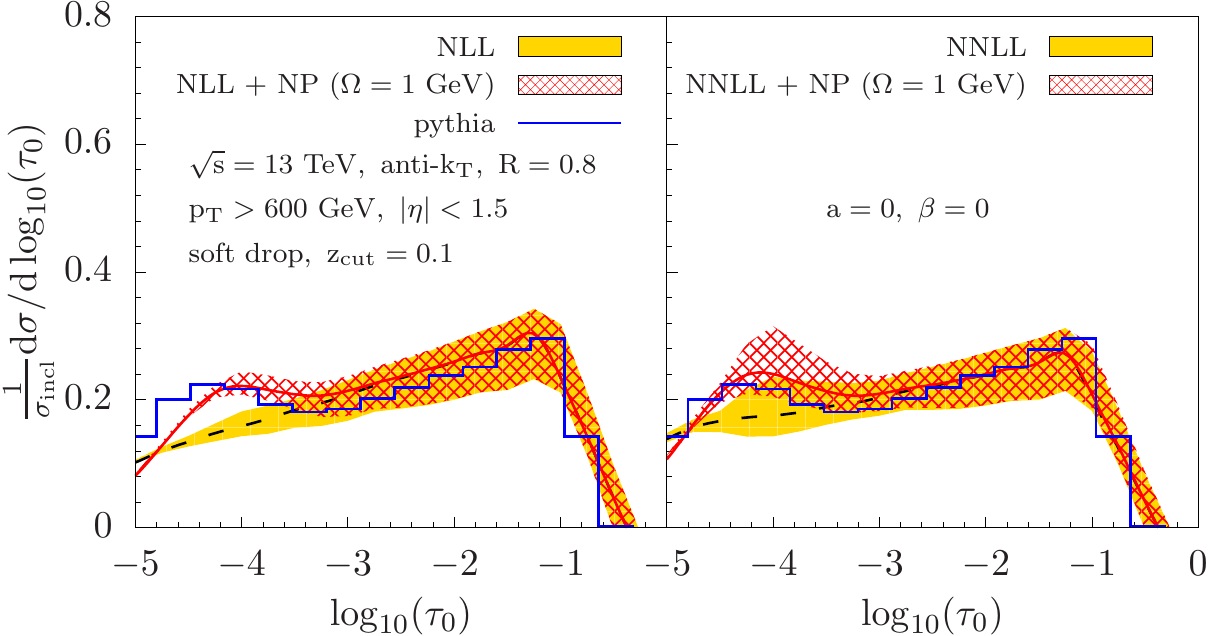}
\vspace*{-.2cm}
\caption{Comparison of the groomed jet mass distribution for $\beta=0$ at NLL (left) and NNLL (right). We use the same kinematical setup as in Fig.~\ref{fig:angularity1} and Fig.~\ref{fig:angularity2} above. \label{fig:angularity3}}
\vspace*{-.2cm}
\end{figure}

We start by presenting results for the groomed angularities at NLL accuracy for LHC kinematics at $\sqrt{s}=13$~TeV. The inclusive jet sample is reconstructed with $R=0.8$ using the anti-k$_T$ algorithm~\cite{Cacciari:2008gp} and we require $p_T>600$~GeV and $|\eta|<1.5$. In Fig.~\ref{fig:angularity1} we show the results for $a=0.5$ (upper panels) and $a=0$ (lower panels) for the grooming parameters $z_{\rm cut}=0.1$ and $\beta=0,1,2$ (left to right) and we normalize our results by the inclusive jet cross section. Fig.~\ref{fig:angularity2} displays the jet angularities for $a=-0.5$ (upper panels) and $a=-1$ (lower panels) for the same kinematical setup. We show the purely perturbative results (yellow bands) obtained from the factorization formula in Eq.~(\ref{eq:factorization}). We obtain the largest QCD scale uncertainties for $a=0.5$. This is expected as the parameter $a$ controls the sensitivity to soft physics, see Eq.~(\ref{eq:jetangularity}). Note that power corrections associated with the soft recoil become increasingly important as $a \to 1$. This problem could be handled conveniently by using a recoil-free axis such as the winner-take-all axis instead of the standard jet axis used here. See for example~\cite{Larkoski:2014uqa}. The corresponding Pythia8 results are shown in blue in Figs.~\ref{fig:angularity1} and~\ref{fig:angularity2}. We show the particle level results from Pythia including initial state radiation (ISR), multi parton interactions (MPI) and hadronization. In order to correct the purely perturbative results to the particle level we include NP effects by convolving with the NP shape function as shown in Eq.~(\ref{eq:convolutionNP}). For the parameter $\Omega_a$ that appears in the NP shape function in Eq.~(\ref{eq:Fk}), we factor out the $a$ dependence as
\be
\Omega_a=\f{\Omega_{a=0}}{1-a} \,,
\ee
which was introduced for angularity measurements in $e^+e^-$ collisions~\cite{Lee:2006nr}. In principle, the NP shape function depends on the grooming parameters $z_{\rm cut}$, $\beta$ and for $\beta\to\infty$ or $z_{\rm cut}\to 0$ the ungroomed results needs to be recovered. A more rigorous field theoretic treatment of the NP shape functions including the dependence on the grooming parameters can be found in~\cite{NPgroom}. Here we choose $\Omega_{a=0}=1$~GeV which, overall, gives a reasonable estimate of the relevant NP physics as shown by the red hatched band. In practice, the value of $\Omega_{a=0}$ can be determined via a global fit to the jet angularity data with different choices of $a$. This way of determining the NP model is one of the major advantages of studying jet angularity distributions.

In Fig.~\ref{fig:angularity3}, we show a comparison of the jet mass distribution $\tau_0$ for $\beta=0$ at NLL (left) and NNLL (right) using the same NP model. We observe that the central values of the NLL and NNLL predictions are very close to each other, which implies the good convergence of the perturbative series. The NNLL central value agrees slightly better with
the Pythia simulations than the NLL one both in the perturbative and non-perturbative regions.
However, we noticed an enhanced QCD scale uncertainty at NNLL in the small $\tau_0$ region, mainly due to the inclusion of the non-zero non-cusp soft anomalous dimensions starting at two-loop. In the future, one may include the two-loop Wilson coefficients of each function to achieve full NNLL$'$ accuracy and to further reduce the theoretical uncertainty. The full NNLO results are either known~\cite{Becher:2006qw,Becher:2010pd} or can be obtained using the existing analytic~\cite{Kelley:2011ng} or numerical techniques~\cite{Boughezal:2015eha}. 

\section{Conclusions \label{sec:4}}

In this work, we studied soft drop groomed jet angularities for inclusive jet production $pp\to{\rm jet}+X$ at the LHC. We presented a factorized form of the cross section in the phenomenologically relevant region of small jet angularities $\tau_a$. The analytical calculation of all relevant ingredients was performed at next-to-leading order. Using renormalization group techniques, the resummation of all relevant large logarithms in the jet radius parameter $R$, the jet angularity $\tau_a$ and the soft drop grooming parameter $z_{\rm cut}$ was achieved. The soft drop grooming procedure makes the jet angularities robust in the complicated LHC environment which allows for a one-to-one comparison between data and first principles calculations in QCD. We presented numerical results for representative LHC kinematics at $\sqrt{s}=13$~TeV for different values of the jet angularity parameter $a$ and the grooming parameter $\beta$. In order to estimate the impact of non-perturbative physics, we compared to Pythia results at the particle level. Overall we found good agreement after including non-perturbative effects through a shape function. We expect that our results will be very relevant for future extractions of the QCD strong coupling constant $\alpha_s$ from jet substructure data taken at the LHC. From the theory side it is crucial to extend the theoretical accuracy beyond NLL in order to achieve a competitive extraction of $\as$. In this work, we performed a step in this direction by extending the resummation of all relevant large logarithms to NNLL accuracy for the case of the jet mass distribution and $\beta=0$ for which we found an improved agreement with the results from Pythia. In the future, the more general cases can be obtained using existing numerical techniques. In addition, the matching with the full NNLO calculations for inclusive jet production~\cite{Currie:2016bfm} need to be carried out in order to meet the precision requirements for a reliable determination of $\alpha_s$. We also expect that the measurement of (groomed) jet angularities for different values of $a$ in heavy-ion collisions will allow for a more complete understanding of how jets get modified as they traverse the dense and hot QCD medium.

\medskip

{\bf Acknowledgments}

We thank D.~d'Enterria, S.~Hoeche, Y.-J. Lee and B.~Nachman for inspiring and helpful discussions. Z.K. is supported by the National Science Foundation under Grant No.~PHY-1720486. K.L. is supported by the National Science Foundation under Grants No.~PHY-1316617 and No.~PHY-1620628. X.L. is supported by the National Natural Science Foundation of China under Grant No.~11775023 and the Fundamental Research Funds for the Central Universities. F.R. is supported by the Department of Energy under Contract No.~DE-AC0205CH11231, and the LDRD Program of Lawrence Berkeley National Laboratory.

\end{document}